\documentclass[sigconf,nonacm]{acmart}

\usepackage{balance}
\AtBeginDocument{%
  \providecommand\BibTeX{{%
    \normalfont B\kern-0.5em{\scshape i\kern-0.25em b}\kern-0.8em\TeX}}}

\renewcommand\footnotetextcopyrightpermission[1]{}

\begin{document}

\title{Multi-Objective Recommender Systems: Survey and Challenges}
\titlenote{Proceedings of the 2nd Workshop on Multi-Objective Recommender Systems (MORS) at 16th ACM Conference on Recommender Systems (RecSys), 2022, Seattle, USA.}
\author{Dietmar Jannach}
\email{dietmar.jannach@aau.at}
\affiliation{%
  \institution{University of Klagenfurt}
  \country{Austria}
}
\begin{abstract}
Recommender systems can be characterized as software solutions that provide users convenient access to relevant content. Traditionally, recommender systems research predominantly focuses on developing machine learning algorithms that aim to predict which content is relevant for individual users. In real-world applications, however, optimizing the accuracy of such relevance predictions as a single objective in many cases is not sufficient. Instead, multiple and often competing objectives have to be considered, leading to a need for more research in multi-objective recommender systems. We can differentiate between several types of such competing goals, including \emph{(i)} competing recommendation quality objectives at the individual and aggregate level, \emph{(ii)} competing objectives of different involved stakeholders, \emph{(iii)} long-term vs.~short-term objectives, \emph{(iv)} objectives at the user interface level, and \emph{(v)} system level objectives. In this paper we review these types of multi-objective recommendation settings and outline open challenges in this area.
\end{abstract}

\keywords{Recommender systems, Multi-objective optimization, Multistakeholder recommendation, Short-term and long-term objectives}
\maketitle

\section{Introduction}
\label{sec:introduction}
Generically defined, recommender systems can be characterized as \emph{software solutions that provide users convenient access to relevant content}. The types of conveniences that such systems provide can be manifold. Historically, recommender systems were mainly designed as information filtering tools, like the early GroupLens system~\cite{ResnickGrouplens1994} from 1994.
Later on, various other ways were investigated how such systems can create value, e.g., by helping users \emph{discover} relevant content,
by providing easy access to related content (e.g., accessories),
or by even taking automatic action like creating and starting a music playlist.

While a recommender systems can serve various purposes and create value in different ways \cite{Jannach2021SessionRSHBImpact}, the predominant (implicit) objective of recommender systems in literature today can be described as ``guide users to relevant items in situations of information overload'', or simply ``find good items''~\cite{Herlocker:2000:ECF:358916.358995}. The most common way of operationalizing this information filtering problem is to frame the recommendation task as a supervised machine learning problem. The core of this problem is to learn a function from noisy data, which accurately predicts the \emph{relevance} of a given item for individual users, sometimes also taking contextual factors into account.

Although the actual relevance of recommended items can be assessed in different ways~\cite{GunawardanaS15}, data-based offline experiments dominate the research landscape. In early years, rating prediction was considered a central task of a recommender and the corresponding objective was to minimize the mean absolute error (MAE), see~\cite{Shardanand1995} for work using MAE in 1996. Nowadays, item ranking is mostly considered to be more important than rating prediction, and a variety of corresponding ranking accuracy measures are used today.

While the metrics changed over time, the research community has been working on optimizing relevance predictions in increasingly sophisticated ways for almost 30 years now. The main objective of such research is to minimize the relevance prediction error or to maximize the accuracy of the recommendations. The underlying assumption of these research approaches is that better relevance predictions lead to systems that are more valuable for their users. This seems intuitive for many practical applications, because a better algorithm should surface more relevant items in the top-N lists shown to users.

Such an assumption might however not always be true, and it was pointed out many years ago that ``being accurate is not enough''~\cite{McNee:2006:AEA:1125451.1125659}. A recommender system might for example present users with obvious recommendations, e.g., recommending new Star Wars sequels to a Star Wars lover. The prediction error for such recommendations might be even close to zero. But so will be the value of the recommendations to users, who most probably know these movies already. Observations like this led to a multitude of research efforts on ``beyond-accuracy'' measures like diversity, novelty, or serendipity, see~\cite{Bradley-AICS-2001} for an early work 
from 2001. Which beyond-accuracy dimension is relevant for a given setting depends on the purpose the recommender is intended to serve~\cite{JannachAdomavicius2016purpose}. Independent of the purpose, beyond-accuracy measures however often compete with accuracy measures, leading to the problem that multiple objectives have to be balanced when serving recommendations.

Historically, when considering the purpose of a recommender system, the focus of research was on the value of such a system for \emph{consumers}. Only in recent years more attention was paid to the fact that recommender systems in practice factually serve some business or organizational objectives. Considering these provider-side aspects therefore requires that we see recommendation as a problem where the interests and objectives of multiple stakeholders must be considered \cite{Abdollahpouri2022,abdollahpouri2020}, often also taking different optimization time horizons into account.

Overall, while being able to predict the relevance of individual items for users remains to be a central and relevant problem, considering only one type of objectives, i.e., prediction accuracy, and the corresponding metrics may be too simplistic and ultimately limit the impact of academic research efforts in practice. One important way to escape the limitations of current research practice is to consider multiple quality dimensions and stakeholder objectives in parallel~\cite{jannach2021mcnamara}. Next, in Section~\ref{sec:taxonomy}, we will discuss various forms of multi-objective recommender systems found in the literature.

\section{Types of Multi-Objective Recommendation Settings}
\label{sec:taxonomy}
\begin{figure*}[h!t]
\centering
\includegraphics[scale=0.50]{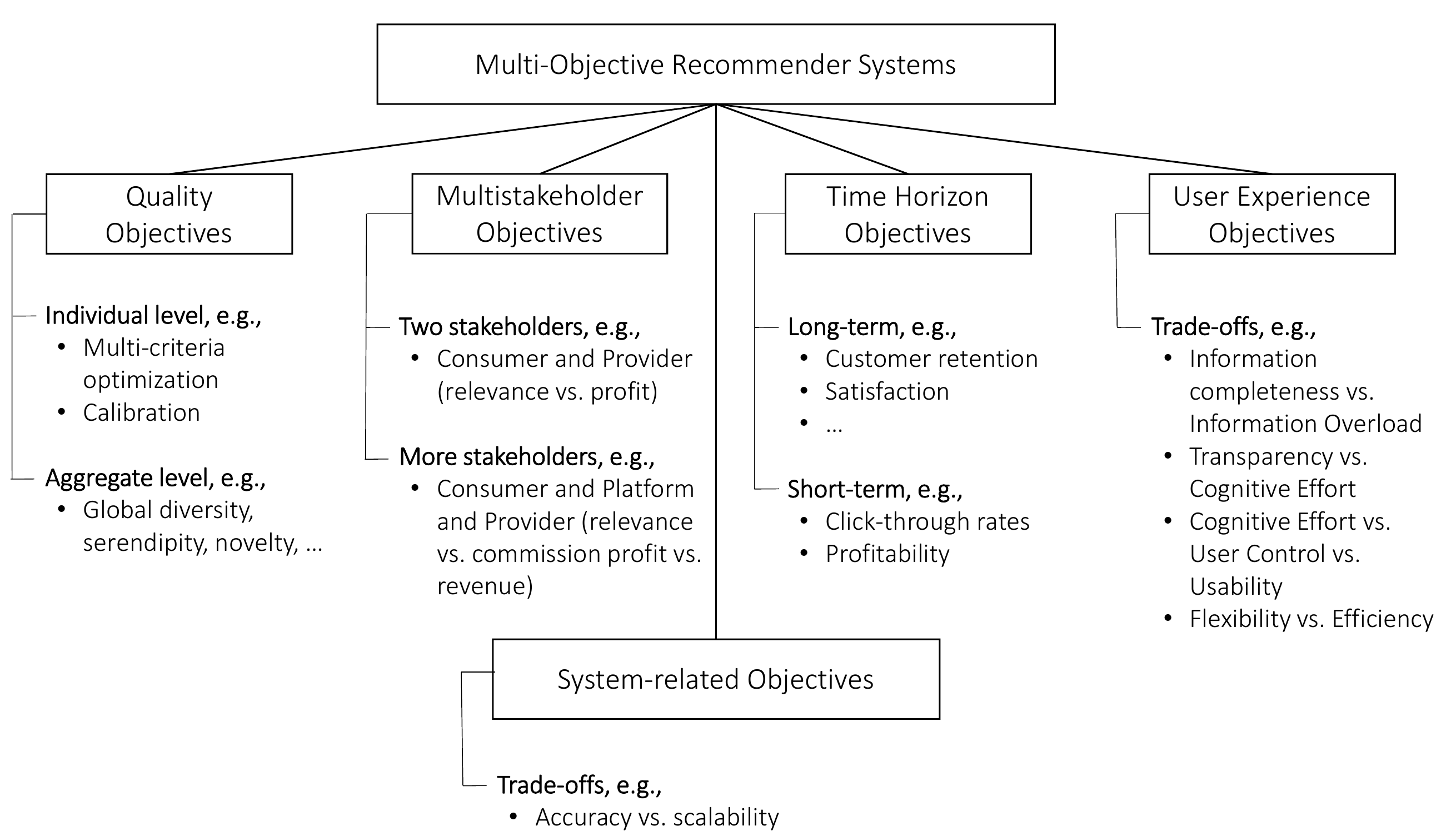}
\caption{Taxonomy of different types of multi-objective recommendation settings.
}
\label{fig:taxonomy}
\end{figure*}

On a very general level, we can define that ``\emph{a multi-objective recommender system (MORS) is a system designed to jointly optimize or balance more than one optimization goal.}'' Figure~\ref{fig:taxonomy} provides an taxonomy of different and mostly orthogonal types of multi-objective recommendation settings. We note that the various objectives to be pursued with a recommender system are in many cases competing. Diversity goals, for instance, typically stand in contrast with accuracy. However, not all objectives necessarily represent such a contrast. Increasing the musical coherence of a playlist for example showed to be advantageous in terms of accuracy in some cases in~\cite{JannachLercheEtAl2015b}. Also, when considering short-term and long-term objectives, taking measures to increase interactivity and engagement with the system in the short term is usually considered beneficial for customer retention in the long run~\cite{Gomez-Uribe:2015:NRS:2869770.2843948}. Note that while most discussed objectives are \emph{algorithmic} ones, there may also be \emph{non-algorithmic} objectives at the user experience level, i.e., objectives that are not tied to a specific underlying algorithm.

\subsection{Recommendation Quality Objectives}
\label{subsec:quality-objectives}
Under this category, we subsume problem settings, where more than one quality objective of recommendations for users must be considered. We can differentiate between the system considering such objectives at the level of individual users or at an aggregate level, i.e., for the entire user base.

\paragraph{Individual level.} At the individual level, consumers can have specific (short-term) preferences, e.g., regarding item features that should be considered in parallel. For instance, a user of a hotel booking platform might be interested in a relatively cheap hotel, which in addition is close to the city center. In such a situation, a recommender system might therefore strive to find an offering that respects both preferences as good as possible. Such problem settings often occur in \emph{interactive} or \emph{conversational} recommendation scenarios. Technically, a variety of Multi-Criteria Decision-Making (MCDM) methods such as Analytic Hierarchical Process (AHP), the Weighted Sum Method (WSM) or Multi-Attribute Utility Theory (MAUT) can be applied, see~\cite{Triantaphyllou2000,DBLP:journals/www/ManouselisC07}. Such methods can also be part of interactive constraint-based recommendation approaches~\cite{Jannach:2004:ASK:3000001.3000153} to rank items. In addition, some of these constraint-based approaches support the automatic relaxation of individual user preferences in case none of the remaining items matches all consumer preferences~\cite{FelfernigFriedrichEtAl2006}.

A different way to consider multiple user preferences at the individual level is called \emph{calibration}. In such an approach, the idea usually is to match certain beyond-accuracy aspects of the recommendation list with past preference profiles of individual users. In an early work, Oh et al.~\cite{Oh2011} tried to align the recommendations with the past popularity tendencies of a user. Later, Jugovac et al.~\cite{JugovacJannachLerche2017eswa} extended the approach for multiple optimization objectives. A more formal characterization of calibration was introduced in \cite{steck2018calibration} by Steck, and \cite{DBLP:conf/um/AbdollahpouriMB21} represents another recent work into that direction. In most cases the central idea of these approaches  is to match two distributions, e.g., the popularity distribution of items in the user profile and the popularity distribution of recommendations. An alternative optimization goal was used in \cite{JannachLercheEtAl2015b} for the music domain, where the objective was to find musically coherent playlist continuations while preserving prediction accuracy.

\paragraph{Aggregate level.}
The majority of published research on balancing different recommendation quality aspects targets the aggregate level. The objective of such works is to balance the recommendations for the entire user base, the corresponding metrics are therefore usually averages. The most common beyond-accuracy measures in the literature include diversity, novelty, serendipity, catalog coverage, popularity bias or fairness, see, e.g., \cite{DBLP:journals/tkde/AdomaviciusK12,Kaminskas:2016:DSN:3028254.2926720,Vargas:2011:RRN:2043932.2043955,Abdollahpouri2017Controlling,Fairness2022Ekstrand}. Most commonly, the goal is to balance accuracy with exactly one of these measures, assuming that there is a trade-off between these quality factors. Increasing diversity is for example commonly assumed having a negative impact on accuracy metrics. A few works exist which consider more than two factors. In an earlier work in this area \cite{Rodriguez2012Multiple}, the authors describe an effort to build a \emph{talent recommendation} system at LinkedIn, which not only considers the semantic match between a candidate profile and a job, but which also take side constraints into account, for instance, the presumed willingness of a candidate to change positions.

Technically, a variety of approaches to balance competing goals can be found in the literature. Re-ranking accuracy-optimized lists is probably the most common approach and was also used in early approaches for diversification \cite{Bradley-AICS-2001}. Alternative techniques for creating a balanced lists or for merging different lists include constraint optimization techniques, graph-based and neighborhood-based approaches, methods that are based on the concept of Pareto efficiency, methods that integrate different objectives in their loss function, or bandit-based approaches that address the explore-exploit problem for improved discovery through novel recommendations  \cite{Zhang2008,Jambor2010,Zhang2020Auralist,said2013,Ribeiro2015,gabrieljannach2019,McInerney2018,zhou2020,ISUFI2021102459}.

\subsection{Multistakeholder Objectives}
The beyond-accuracy quality metrics discussed in the previous section were historically mostly introduced to improve recommendations for end users. Higher diversity, for example, should avoid monotonicity, and novelty should support discovery. The underlying assumption---also of pure accuracy-oriented works---is that improving different quality aspects for users would be at least indirectly beneficial for the providers. Only in recent years, more attention was paid in the literature to the fact that many recommendation scenarios in the real world are situated in environments, where the objectives of multiple stakeholders have to be considered. The common players in such \emph{multistakeholder} recommendation problems include \emph{end consumers}, recommendation \emph{service providers}, item \emph{suppliers}, and sometimes even parts of a broader \emph{society} \cite{abdollahpouri2020,jannach2021mcnamara}. In such settings, a recommender system may serve different purposes for different stakeholders \cite{JannachAdomavicius2016purpose}, and the related objectives may stand in conflict.\footnote{In some cases there may even be subgroups within the consumer stakeholder group, e.g., free vs.~premium or new vs.~existing customers, for which different objectives may exist.}

A typical problem setting in practice that involves \emph{two stakeholders} is that of balancing consumer and provider objectives. In many cases, there may be a potential trade-off between (a) recommending the \emph{most relevant} items for consumers and (b) recommending items that are also somewhat relevant but assumed to be favorable in terms of the provider's business objectives. Some of the discussed beyond-accuracy metrics can actually be seen as serving both stakeholders. Making more novel recommendations not only potentially leads to a better user experience, but also to more engagement with the service and longer term customer retention \cite{Anderson2020}.

A number of research works however also consider business (or: organizational) objectives more directly, in particular in the form of recommender systems that are ```price and profit aware'' \cite{DBLP:journals/corr/JannachA17}. In \cite{CHEN20081032}, for instance, two heuristic profit-aware strategies are proposed and the authors found that such methods can increase the profit from cross-selling without losing much recommendation accuracy. A number of other possible solutions to maximize provider profitability were proposed over the years, using a variety of approaches including mathematical optimization with side constraints, re-ranking accuracy-optimized lists based on profit considerations, random walks over graphs or analytical modeling \cite{Azaria2013,WANG20097299,Azaria2013,Panniello2016}

Besides such situations with potential consumer-provider trade-offs, there are settings that involve even more stakeholders to consider. A problem that has been studied for several years---even though not under the name multistakeholder recommendation---is that of \emph{group recommendation} \cite{Masthoff2015GroupRS}. In such settings, the system's goal is to determine a set of recommendations that suit the preferences of a group of users, e.g., friends who want to watch a movie together. A variety of strategies to aggregate the individual user preferences were proposed over the years. Early works on the topic can be found in \cite{OConnor2001} and \cite{Masthoff2004}. In \cite{Masthoff2004}, for instance, Masthoff reports the outcomes of different user studies aimed to understand how humans make choices for a group and finds that humans indeed sometimes follow strategies inspired from Social Choice Theory. We note here that the group recommendation setting differs from other multistakeholder scenarios in that all stakeholders receive the same set of recommendations.

\emph{Reciprocal recommendation} is another specific set of problem settings involving multiple stakeholders. Here, instead of recommending items to users, the problem is to recommend users to users, also known as people-to-people recommendation. Typical application scenarios are recommendations on dating and recruiting platforms. A particularity of such settings is that the success of a recommendation is not determined solely by the recipient of the recommendation, but there must be a mutual preference match or compatibility between the two people involved, see ~\cite{PALOMARES2021103} for an in-depth discussion on the topic. The recommendation service provider therefore faces additional complexities in the matching process and in parallel has to observe its own business objectives and constraints. On a job recommendation platform, for example, the provider may have to additionally ensure that each paid job advertisement receives a minimum number of relevant impressions, i.e., exposure \cite{DBLP:conf/recsys/AbelDEK17}.

Similar considerations may generally apply when the recommendation platform serves as a marketplace with multiple suppliers of identical or comparable items. Let us consider again the example of a typical hotel booking platform, which serves personalized recommendations to its users \cite{jannach2021mcnamara}. Besides the consumer, who already might have competing objectives, there are the property owners, who have their offerings listed on the booking platform and pay a commission for each booking. The goal of the property owners is that their offerings are exposed to as many matching customers as possible in order to increase the chances of being booked.
The booking platform, finally, is not only interested in recommending matching hotels to consumers, but might also seek to maximize their commission, e.g., by recommending slightly more expensive hotels. In addition, to balancing these objectives, the platform may furthermore have to ensure that \emph{all} listed properties reach a sufficient level of exposure, i.e., chance of being booked. This may be required to ensure a long-term relationship with property owners, who might otherwise discontinue to list their offerings on the platform at some stage \cite{Krasnodebski2016considering}. We discuss short-term and long-term objectives next.

\subsection{Time Horizon Objectives}
\label{subsec:time-horizon-objectives}
In some application domains, it might be quite simple to increase short-term Key Performance Indicators (KPIs). In the hotel booking scenario which we have just discussed, boosting short-term revenue might be achieved by recommending hotels with currently discounted rates, which maximizes the probability of a transaction \cite{Jannach2017Session}. In the news domain, recommending articles on trending topics, articles with click-bait headlines or generally popular content such as celebrity gossip
may lead to high click-through-rates (CTR). In the music domain, recommending tracks of trending or popular artists, which the user already knows, may be a safe strategy when the target metric is to avoid ``skip'' events.

Such strategies that are successful in the short term may however be non-optimal or even detrimental in the long run. The recommendation of discounted hotel rooms may be bad for profit. News readers may be disappointed when actually reading articles with a clickbait headline and may not trust these recommendations in the future. Music listeners finally may have difficulties to discover new artists over time and may quit using the service after some time.

Most academic research is based on one-shot evaluations, typically focusing on the prediction accuracy given a static dataset and a certain point in time. Longitudinal effects of different recommendation strategies are much less explored and there is also limited literature on the long-term effects of recommender systems in industry. A/B tests in industry may last from a few weeks to several months. In \cite{Gomez-Uribe:2015:NRS:2869770.2843948}, the case of Netflix is discussed, where one main KPI is customer retention, which is oriented towards the long-term perspective. In their case, attributing changes in the recommender system to such long-term effects is reported to be challenging, e.g., because of already high retention rates and the need for large user samples. Other reports from real-world deployments a recommender systems can be found in \cite{Panniello2016} or \cite{LeeHosanagar2018}. In \cite{LeeHosanagar2018}, the authors for example found that using a recommender system led to decreased sales diversity compared to a situation without a recommender. A similar effect was reported in \cite{Anderson2020}, where the recommender system on a music streaming site led to a reduced consumption diversity. A survey of other reports on real-world applications of recommender systems can be found in \cite{jannachjugovactmis2019}.

Given the limitations of one-shot evaluations, we have observed an increased interest in longitudinal studies in recent years. One prominent line of research lies in the area of reinforcement learning (RL) approaches in particular in the form of contextual bandits, see e.g.~\cite{Li2010Contextual} for an earlier work in the news domain. In such approaches, the system sequentially selects items to recommend to users and then incorporates the users' feedback for subsequent recommendations. Different recommendation algorithms can be evaluated offline with the help of simulators, e.g., \cite{rohde2018recogym,McInerney2021Accordion}. A common challenge in this context is to ensure that such evaluations are unbiased \cite{Li2011Unbiased,Huang2020}. We note that the consideration of temporal aspects such as different time horizons or delayed feedback have been explored in the RL literature for the related problem of computational advertising for several years \cite{Theocharous2015,ModelingDelayed2014}.

Reinforcement learning approaches typically aim at finding a strategy to maximize the expected \emph{reward}. During the last few years, a number of studies that use other forms of simulations were published that focus on other important long-term phenomena of recommender systems. These studies for example focus on longitudinal effects of recommender systems on sales diversity \cite{Fleder:2009:BCN:1538966.1538968}, potential reinforcement effects in terms of popularity bias and other aspects for traditional and session-based recommendations \cite{Jannach2015umuaiWhat,ferrarojannachserra2020recsys}, longitudinal performance effects of recommender systems and the ``performance paradox'' \cite{Gedas2019Longitudinal}, differences in terms of long-term effects of consumer-oriented and profit-oriented recommendation strategies \cite{ghanem2022simulation}.

\subsection{User Experience Objectives}
\label{subsec:user-experience-objectives}
Going beyond specifics of individual algorithms, there can be also various objectives to be pursued at the user interaction level of a recommender system. The design space for the user interface of recommender systems is actually large, see \cite{JugovacJannachTiis2017}, and there thus may be a number of competing objectives at the user interface (UI) level.

Here, we only list a few examples of potential trade-offs that may be common for many recommender system applications.
\begin{itemize}
  \item \emph{Information Completeness vs.~Information Overload}: This, for instance, refers to the question of how many items should be shown to users and if we should completely filter out certain items from the result list. Showing too few options may give users the feeling that the system holds back some information. If there is too much information users will find themselves again in a situation of information overload \cite{DBLP:conf/recsys/BollenKWG10,aljukhadar2012using}. Besides the question of how many options to show, a related question is how much detail and additional information to show for each recommendation.
  \item \emph{Transparency and User Control vs.~Cognitive Effort}: Transparency and explanations are commonly considered to be trust-establishing factors in recommender systems~\cite{Pu:2011:UEF:2043932.2043962}. A variety of different ways of explaining recommendations were proposed in the literature \cite{tintarev2011designing,Explanations:UMUAI2017}. Many of these academic proposals are quite complex and may easily cognitively overload average end users. Similar considerations apply for approaches that implement mechanisms for \emph{user control} in recommender systems \cite{Ekstrand2015,JannachNaveedEtAl2016}.
  \item \emph{Flexibility vs.~Efficiency}: This question arises in the context of modern conversational recommender systems that are implemented in the form of chatbots. Chatbots typically support two forms of interactions: a) natural language input and b) form-based input (i.e., using buttons). While natural language inputs may allow for more flexible interactions, the study in \cite{Iovine2020Conversational}, for instance, indicated that a combination of interaction modalities was most effective.

\end{itemize}
Several other more general design trade-offs may exist depending on the specific application, e.g., regarding acceptable levels automating adaptivity of the user interface, which may hamper usability~\cite{Paymans2004}.

\subsection{System Objectives}
\label{subsec:system-objectives}
In this final category we discuss technical aspects and their potential trade-offs. We call them ``system objectives'', as they refer to more general system properties.

One such trade-off in practice may lie in the complexity of the underlying algorithms and the gains that one may obtain in terms of business-related KPIs. Already in the context of the Netflix Prize \cite{bennett2007netflix} we could observe that the winning solutions were finally not put into production, partly due to their complexity. Similar considerations can be made for today's sometimes computationally demanding methods based on deep learning. In some cases there might be a diminishing return of deploying the most sophisticated models in production, only because they lead to slightly better accuracy values in offline testing. In some research works, it even turns out that ``embarrassingly shallow'' models can be highly competitive in offline evaluations \cite{Steck2019Easer}.

With highly complex models, not only scalability issues may arise and monetary costs for computing resources may increase, the complexity of the architectures might also make such systems more difficult to maintain, debug, and explain. On the other hand, solutions built upon modern deep learning frameworks are sometimes reported to be advantageous over conceptually simpler, but specialized solutions, because these frameworks and deep learning architectures make it very easy to integrate various types of information into the models \cite{Steck2021AIMag}.

\section{Summary and Challenges}
\label{sec:discussion}
Our review outlines that providing automated recommendations is a problem that may require the consideration of more than one objective in many real-world use cases. Such multi-objective settings may include competing objectives of consumers, possible tensions between goals of different stakeholders, conflicts when optimizing for different time horizons, competing design choices at the UI level, as well as system-level and engineering-related considerations. In this work, we reviewed the literature in this area and provided a taxonomy to organize the various dimensions of multi-objective recommendation. We note here that the categories of the taxonomy are not mutually exclusive. For instance, a multi-objective recommendation approach may address both aspects regarding different time horizons as well as the possibly competing goals of the involved stakeholders.

In practice, one main challenge may usually lie in deciding on the right balance between the competing goals from an organizational perspective. Various stakeholders from different organizational units may have to agree on such decisions, and corresponding KPIs need to be defined and monitored. Given these KPIs, suitable optimization goals and possibly proxy measures have to be implemented and validated at the technical level.

In academic settings, researchers typically abstract from the specifics of a given application context, aimed at developing generalizable algorithmic solutions to deal with multi-objective problem settings. This abstraction process commonly involves the use of \emph{offline evaluation approaches}, the establishment of certain assumptions, and the introduction of computational metrics which should be optimized. After such an abstraction, one main challenge, however, lies in the evaluation process and, in particular, in making sure that improvements that are observed in terms of abstract evaluation measures would translate to better systems in practice~\cite{cremonesi2021aimag}.

Unfortunately, in many of today's research works, we observe phenomena similar to the ``abstraction traps'' described by Selbst et al.~\cite{Selbst2019Fairness} in the context of research on algorithmic works in \emph{Fair Machine Learning}. In the case of competing individual-level quality goals, for example, how can we be sure that a particular diversity metric, which we optimize such as an intra-list similarity, matches human perceptions and what would be the right balance for a given application setting or an individual user? How do we know if calibrated recommendations are liked more by users, and what would be the effects of calibration on organizational goals? Answering such questions requires corresponding user studies to, e.g., validate that the computational metrics are good proxies for human perceptions.

The problem however becomes even more challenging when not even the target concepts are entirely clear. In recent years, a widely investigated multi-objective problem setting is the provision of \emph{fair} recommendations \cite{Fairness2022Ekstrand}. Unfortunately, optimizing for fairness turns out to be challenging, as fairness is a societal construct. Researchers therefore came up with various types of ways of operationalizing fairness constraints. However, in many of such works, little or no evidence or argumentation is provided why the chosen fairness metrics are meaningful in practice in general or in a particular application setting, see \cite{Deldjoo2022FairnessSurvey} for a survey on the recent literature. In several cases, making fair recommendations is simply equated with reducing the popularity bias of recommendations, e.g., by matching it with a target distribution, which is assumed to be given. In reality, however, it is not clear if it would be fair to frequently recommend items that are not popular---these items might simply be of poor quality---or if users would even \emph{perceive} these recommendations to be fair.\footnote{Avoiding popularity biases in general, i.e., beyond fairness considerations, certainly is important for different reasons in practice, e.g., to ensure a certain level of exposure for novel (cold-start) items. } Similar considerations apply for many other types of multi-objective recommender systems.

Overall, these observations call for more studies involving humans in the evaluation loop and industry partners in the research process. However, only few works exist in that direction so far. An example of a user study can be found in \cite{Azaria2013}, outcomes of field studies are described in \cite{Panniello2016}. An offline evaluation with real-world data from industry is done in~\cite{Mehrotra2018Towards}, but even in this case it is not clear if the computational metrics truly correspond to the real-world goals, e.g., if more listening events on the music platform lead to higher user satisfaction as claimed. In a very recent work, Wang and colleagues from Google~\cite{Wang2022Surrogate} investigated the relationship between observed user behavior---using data from a large platform---and desired long-term outcomes in terms of user experience. With their work, they aim at providing guidance for researchers and practitioners when selecting surrogate measures to address the difficult problem of optimizing for long-term objectives.

Ultimately, despite such recent progress, multi-objective recommender systems remains a highly important research area with a number of challenging research questions. Addressing such questions will however help to pave the way towards more impactful recommender systems research in the future.

\balance
\bibliographystyle{ACM-Reference-Format}
\bibliography{main-arxiv}

\end{document}